\magnification \magstep1
\input amssym.def
\input amssym.tex
\def \sf{\Sigma_F}
\def \si{\Sigma_I}
\def \ep{\epsilon}
\def \lam{\lambda}
\def \ni{\noindent}
\def \rta{\rightarrow}
\def \ms{\medskip}

\def \in{\indent}
\def \ss{\smallskip}
\bigskip
\centerline{\bf Spin-$1\over 2$ Amplitudes in Black-Hole Evaporation}
\bigskip
\centerline{ A.N.St.J.Farley and P.D.D'Eath }
\bigskip
\smallskip
\centerline{Department of Applied Mathematics and Theoretical Physics,
Centre for Mathematical Sciences,} 
\smallskip
\centerline{University of Cambridge, Wilberforce  Road, Cambridge CB3 0WA,
United Kingdom}
\bigskip
\centerline{Abstract}
\smallskip
\noindent
In recent papers, we have studied the quantum-mechanical decay of a
Schwarzschild-like black hole, formed by gravitational collapse, into
almost-flat space-time and weak radiation at a very late time.  In
this recent work, we have been concerned with evaluating quantum 
amplitudes (not just probabilities) for transitions from initial 
to final states.  In a general asymptotically-flat context, 
one may specify a quantum amplitude by posing boundary data on (say) 
an initial space-like hypersurface ${\,}\Sigma_I$ and a final space-like 
hypersurface $\Sigma_F{\,}$.  To complete the specification, 
one must also give the Lorentzian proper-time interval between 
the two boundary surfaces, as measured near spatial infinity.  
We have assumed that the Lagrangian contains Einstein gravity 
coupled to a massless scalar field $\phi{\,}$, plus possible 
additional fields;  there is taken to be a 'background'
spherically-symmetric solution $(\gamma_{\mu\nu}{\,},\Phi)$ of the
classical Einstein/scalar field equations.  For bosonic fields, the
gravitational and scalar boundary data can be taken to be $g_{ij}$ 
and $\phi$ on the two hypersurfaces, where $g_{ij}{\;}(i,j=1,2,3)$ 
gives the intrinsic 3-metric on the boundary, and the 4-metric is
${\,}g_{\mu\nu}{\;}{\,}(\mu,\nu =0,1,2,3)$, the boundary being taken
locally in the form $\{x^{0}={\rm const.}\}$.  The classical 
boundary-value problem, corresponding to the calculation of this 
quantum amplitude, is badly posed, being a boundary-value problem 
for a wave-like (hyperbolic) set of equations.  Following Feynman's 
$+i\epsilon$ prescription, one makes the problem well-posed by 
rotating the asymptotic time-interval $T$ into the complex:
$T\rightarrow{\mid}T{\mid}\exp(-i\theta)$, with 
${\,}0<\theta\leq\pi/2{\,}$.  After calculating the amplitude 
for ${\,}\theta >0{\,}$, one then takes the 'Lorentzian limit' 
$\theta\rightarrow 0_{+}{\,}$.  Such quantum amplitudes have 
been calculated for weak ${\,}s=0{\,}$ (scalar), 
${\,}s=1{\,}$ (photon) and ${\,}s=2{\,}$ (graviton) anisotropic 
final data, propagating on the approximately Vaidya-like background
geometry, in the region containing radially-outgoing black-hole
radiation.  In this paper, we treat quantum amplitudes for the case of 
fermionic massless spin-${{1}\over{2}}$ (neutrino) final boundary data.  
Making use of boundary conditions originally developed for local 
supersymmetry, we find that this fermionic case can be treated
in a way which parallels the bosonic case.  In particular, we
calculate the classical action as a functional of the fermionic data 
on the late-time surface $\Sigma_F{\,}$; the quantum amplitude 
follows straightforwardly from this.\par
\medskip
\noindent
{\bf 1. Introduction}
\medskip
\indent
This paper treats quantum amplitudes involving neutrinos 
$(s={{1}\over{2}})$ in the context of nearly-spherical 
Einstein/massless-scalar gravitational collapse to a black hole, 
following the approach of [1-9].  We first summarise the main 
ideas at the bosonic level.  Writing 
${\,}g_{\mu\nu}{\;}(\mu,\nu = 0,1,2,3)$ for the 4-metric and 
${\,}h_{ij}=g_{ij}{\;}(i,j=1,2,3)$ for the intrinsic Riemannian
spatial 3-metric on a surface $\{x^{0}={\rm const.}\}$, one poses 
appropriate boundary data $(h_{ij},\phi)_{I,F}$ (say) on an initial 
space-like hypersurface $\si$ at $x^{0}=0$ and on a final surface 
$\sf{\,}$.  For simplicity, we assume that both $\si$ and 
$\sf$ are diffeomorphic to ${\Bbb R}^{3}$.  The large (real) 
Lorentzian time-separation $T$ between the surfaces is measured 
at spatial infinity.  Again, for simplicity, the initial data 
$(h_{ij},\phi)_I$ are taken to be exactly spherically symmetric, 
corresponding to a diffuse, slowly-moving, initially-spherical 
configuration.  The space-time geometry near the final surface 
$\sf$ is taken to be approximately a Vaidya metric [4,5], 
describing the region which contains radially-outgoing radiation 
due to the collapse to a black hole, except for small perturbations 
in the gravitational and scalar data.\par
\ss
\in
We assume that there is a 'background' Lorentzian-signature 
spherically-symmetric solution of the coupled field equations, 
with 4-metric $\gamma_{\mu\nu}$ and scalar field $\Phi{\,}$.  
If the data for the gravitational and scalar fields are perturbed 
slightly from spherical symmetry, the resulting 
non-spherically-symmetric solution can be written in the form 
$$g_{\mu\nu}{\;} 
={\;}\gamma_{\mu\nu}{\,}+{\,}\epsilon{\,}h^{(1)}_{\mu\nu}{\,} 
+{\,}\epsilon^{2}{\,}h^{(2)}_{\mu\nu}{\,}+{\,}\ldots{\;},
{\qquad}{\quad}\phi{\;} 
={\;}\Phi{\,}+{\,}\epsilon{\,}\phi^{(1)}{\,} 
+{\,}\epsilon^{2}{\,}\phi^{(2)}{\,}+{\,}\ldots{\;},\eqno(1.1)$$
\noindent
where ${\,}\epsilon{\,}$ is a measure of the size of the perturbation.  
We then ask (in this bosonic case) for the quantum amplitude to go 
from asymptotically-flat initial data $(g_{ij}{\,},{\,}\phi)_I$ 
on the initial hypersurface ${\,}\Sigma_I{\,}$ to corresponding 
final data $(g_{ij}{\,},{\,}\phi)_F$ on the final hypersurface 
${\,}\Sigma_{F}{\,}$.  For a complete specification 
of the boundary data involved in the quantum amplitude, one also 
needs the (Lorentzian) time interval ${\,}T$ between 
$\Sigma_I{\,}$ and $\Sigma_F{\,}$, as measured at spatial infinity.\par
\smallskip
\indent
Naively, one might expect there to be a simple (semi-)classical
analogue of this quantum amplitude, as specified by the above boundary
data.  This would involve finding a classical solution of the coupled
Einstein/scalar field equations, which agrees with the boundary data,
and such that the Lorentzian proper-time separation at spatial
infinity is indeed $T{\,}$.  Unfortunately, this will only be possible 
for very special choices of boundary data, having measure zero among all
possible boundary data.  Equivalently, as is well known [2,10,11], the
boundary-value problem for hyperbolic (wave-like) partial differential
equations is not well posed.\par
\smallskip
\indent
The procedure for understanding our (Lorentzian) quantum amplitude
semi-classically is a little more complicated.  Following Feynman's
$+i\epsilon$ proposal [12], we rotate the Lorentzian time-interval 
$T$ into the complex: 
$T\rightarrow{\mid}T{\mid}\exp(-i\theta)$, with
${\,}0<\theta\leq\pi/2{\,}$.  In the extreme case 
${\,}\theta =\pi/2{\,}$, the Riemannian region with positive-definite 
4-metric $g_{\mu\nu}$ is reached.  In the Riemannian case, the field 
equations are 'elliptic {\it modulo} gauge', and one might reasonably 
expect good existence, uniqueness and analyticity properties for 
classical solutions, under suitable conditions on the boundary data.  
In the intermediate complex case, with ${\,}0<\theta <\pi/2{\,}$, 
the interval $T$ and any classical solution will involve the complex 
numbers non-trivially.  In such 'complexified elliptic' cases, it
frequently turns out that the boundary-value problem is 
{\it strongly elliptic}, up to gauge, as defined and discussed 
in [13].  Strongly elliptic partial differential equations are again
liable, as in the real elliptic case, to have good existence, 
uniqueness and analyticity properties [13].  There is reasonable 
ground for expecting that our classical boundary-value problem will
have an (analytic) complex solution, for ${\,}0<\theta\leq\pi/2{\,}$, 
at least under suitable conditions on the boundary data.\par
\smallskip
\indent
One can then form the (complex) classical action $S_{\rm class}{\,}$, 
for the given real boundary data [here $(g_{ij}{\,},{\,}\phi)$], 
as a function of $\theta{\,}$.  Similarly, provided that the quantum 
field theory under consideration has meaningful quantum amplitudes 
(see below), there will be one-loop, two-loop, $\ldots{\;}$ factors 
$A_{0}{\,},{\,}A_{1}{\,},{\,}A_{2}{\,},{\,}\ldots{\,}$,
again depending on the boundary data and on ${\,}\theta{\,}$.  That is, 
the quantum amplitude will have an asymptotic semi-classical expansion
of the form
$${\rm Amplitude}{\quad}\sim{\quad}(A_{0}{\,}+{\,}\hbar{\,}A_{1}{\,} 
+{\,}\hbar^{2}{\,}A_{2}{\,}+{\,}\ldots{\,}){\;}
\exp\bigl(iS_{\rm class}/\hbar\bigr){\quad}.\eqno(1.2)$$
\noindent
Provided that the frequencies involved in the prescribed boundary data
are well below the Planck scale, one expects to be able to approximate
the quantum amplitude by $({\rm const.})\times\exp(iS_{\rm class}){\,}$, 
in units such that $\hbar =1{\,}$.  For dimensional reasons, any loop
corrections will then be extremely small, compared with the leading
term ${\,}A_{0}{\,}\exp\bigl(iS_{\rm class}\bigr){\,}$.  Finally, 
one recovers the Lorentzian quantum amplitude, which is defined to be 
the limit as ${\,}\theta\rightarrow 0_{+}{\,}$ of the 
$\theta$-dependent quantum amplitude.  Provided that one is away 
from the Planck scale, then it is sufficient to compute 
$\lim_{{\,}\theta\rightarrow 0_{+}}\exp(iS_{\rm class}){\,}$.\par
\smallskip
\indent
The only field theories which can give meaningful quantum amplitudes 
in our case, with the property (1.2), are locally supersymmetric, 
that is, supergravity theories with or without supermatter [11,14-16].
The simplest theory of $N=1$ supergravity coupled to supermatter,
which contains Einstein/massless-scalar theory within its bosonic
part, is given in [17].  For this theory, the scalar field has become
complex, with massless spin-${{1}\over{2}}$ partner, and of course 
there is also a spin-${{3}\over{2}}$ gravitino.  It should be 
understood that, in the discussion above concerning quantum 
amplitudes for purely bosonic boundary data, the quantum amplitudes 
refer in fact to the full locally-supersymmetric theory.\par
\smallskip
\indent
It is then natural to ask for quantum amplitudes involving boundary
data which include non-trivial spin-${{1}\over{2}}$ or 
spin-${{3}\over{2}}$ contributions.  In this paper, we shall compute 
the amplitude for non-zero spin-${{1}\over{2}}$ data given on 
$\Sigma_F{\,}$, assuming that the spin-${{1}\over{2}}$ data on 
$\Sigma_I$ are zero.  In particle language, this would refer to the 
production of neutrino-antineutrino pairs.  The resulting classical 
fermionic fields and action will then be elements of a Grassmann 
algebra, as usual in the holomorphic representation for fermions 
[11,18-20].  (In Lorentzian signature, this point of view has also 
been taken in [21], concerning the Cauchy problem for $N=1$ 
supergravity.)  In the present paper, the spin-${{1}\over{2}}$ 
field is described in terms of 2-component spinor fields [22,23], 
the unprimed spinor field $\phi^{A}(x)$ and the independent primed 
spinor field ${\tilde\phi}^{A'}(x){\,}$, not related  to
$\phi^{A}(x){\,}$ by any conjugation operator.  The fields 
$\phi^{A}(x)$ and ${\tilde\phi}^{A'}(x)$ are taken to be odd 
(anti-commuting) elements of a Grassmann algebra [18-20] -- 
see Sec.2 below.  As discussed further in Sec.3, this is the 
spin-${{1}\over{2}}$ description appropriate to Euclidean signature 
(Riemannian geometry), and it is also the description appropriate 
to the 'Riemannian rotated-into-the-complex' case here, with 
${\,}0<\theta<\pi/2{\,}$, and corresponding to a well-posed classical
boundary-value formulation.  The fermionic action (3.1) below 
is quadratic in spin-${{1}\over{2}}$ fields or their derivatives, 
and so is an even element of the Grassmann algebra.  This gives a 
'field description' for the massless spin-${{1}\over{2}}$ theory in 
our case, which allows a treatment of the fermionic quantum 
amplitude which is roughly in parallel with the previous treatments 
for bosonic fields [5,6,8,9]. The holomorphic representation being 
used here for fermions can straightforwardly be related to a 
(possibly more familiar) description in terms of particle states, 
as is outlined in Sec.2.\par
\smallskip
\indent
The basic fields and the spin-${{1}\over{2}}$ action are discussed 
in Sec.3.  The action has the form 
$S_{{1}\over{2}}{\,}
={\,}S_{{{1}\over{2}}V}{\,}+{\,}S_{{{1}\over{2}}B}{\,}$, where
$S_{{{1}\over{2}}V}$ is the standard volume expression, and 
$S_{{{1}\over{2}}B}$ is a boundary contribution, which depends 
on the particular boundary conditions being adopted.  Natural local 
boundary conditions are taken, related to local supersymmetry [24-26].  
These again allow a parallel between our $s={{1}\over{2}}$ treatment 
and that for other (bosonic) spins.  At a solution of the classical 
massless $s={{1}\over{2}}$ field equations (Weyl equations), the 
volume contribution $S_{{{1}\over{2}}V}$ vanishes, whence
${\,}S_{{{1}\over{2}}\rm class}{\,}
={\,}S_{{{1}\over{2}}B\rm class}{\,}$.\par
\smallskip
\indent
Some basic properties of the Vaidya space-time are summarised in
Sec.4.  After the initial gravitational collapse, a large part of the 
classical Einstein/massless-scalar space-time can be approximated 
by a metric of Vaidya type [4,7,27], with its mass decreasing 
extremely slowly with time.  This describes the region of the 
space-time containing the emitted radiation.  Almost all the 
spin-${{1}\over{2}}$ (or other spin) modes of physical interest 
will evolve adiabatically in this background.\par
\ss
\in
In Sec.5, we review the separation of variables for massless 
spin-${{1}\over{2}}$ fields in a Schwarzschild-like background.  
This leads to a treatment of the classical action 
$S_{{{1}\over{2}}B{\rm class}}$ in which local boundary data 
are specified on the final surface $\Sigma_F{\,}$, given, 
for simplicity (as above), that trivial fermionic data are taken 
on $\Sigma_I{\,}$.  The form of the resulting classical action, 
as a functional of the final boundary data, is compared with that 
for massless spin-0 perturbations, found in [5,6].  The form of
the spin-${{1}\over{2}}$ Lorentzian quantum amplitude can then be 
deduced straightforwardly from the definition 
${\,}\lim_{{\,}\theta\rightarrow 0_{+}}\exp(iS_{\rm class}){\,}$.  
A brief conclusion is contained in Sec.6.\par 
\medskip
\ni
{\bf 2. Holomorphic representation for fermions}
\medskip
\indent
Suppose, at first, that one had a Fermi system with only one degree of
freedom [18,19], rather than a quantum field theory.  In the holomorphic
representation [18-20], one would begin by considering two variables 
$a^*$ and $a{\,}$, which are odd (anti-commuting) elements of a
Grassmann algebra: 
$$a^{*}a+aa^{*}=0{\;};{\qquad}(a^{*})^{2}=0{\;};{\qquad}a^{2}=0{\;}.
\eqno(2.1)$$ 
\noindent
A typical element of the resulting Grassmann algebra 
(over the complex numbers) has the form 
$$f(a^{*},{\,}a){\;}{\,}
={\;}{\,}f_{00}+f_{01}{\,}a+f_{10}{\,}a^{*}+f_{11}{\,}aa^{*}{\quad},
\eqno(2.2)$$
\noindent
where $f_{00}{\,},{\,}f_{01}{\,},{\,}f_{10}{\,}$ and $f_{11}$ 
are complex numbers.  Functions of the form
$$f(a^{*}){\;}{\,}={\;}{\,}f_{0}+f_{1}{\,}a^{*}{\quad},\eqno(2.3)$$
with $f_0$ and $f_1$ being complex numbers, are called 
{\it holomorphic functions};  these describe the state vectors 
of the system and form a two-dimensional space.\par
\smallskip
\indent
The operators ${\bf a^{*}}$ and ${\bf a}{\,}$, acting on state vectors 
given in the form (2.3), are defined by
$${\bf a^{*}}\bigl(f(a^{*})\bigr){\;}{\,}
={\;}{\,}a^{*}{\,}f(a^{*}){\;};
{\qquad}{\bf a}\bigl(f(a^{*})\bigr){\;}{\,}
={\;}{\,}{{d}\over{da^{*}}}{\,}f(a^{*}){\quad},\eqno(2.4)$$
\noindent
where fermionic differentiation is defined by 
$${{d}\over{da^{*}}}\Bigl(f_{0}+f_{1}{\,}a^{*}\Bigr){\;}{\,}
={\;}{\,}f_{1}{\quad}.\eqno(2.5)$$
\noindent
Then, one can verify the standard anti-commutation relations
$${\bf a^* a}{\,}+{\,}{\bf a{\,}a^{*}}{\;}{\,}={\;}{\,}1{\;};
{\qquad}{\qquad}({\bf a^{*}})^{2}{\;}{\,}={\;}{\,}0{\;};
{\qquad}{\qquad}({\bf a})^{2}{\;}{\,}={\;}{\,}0{\quad}.\eqno(2.6)$$
\smallskip
\indent
Berezin integration [18-20,28] is defined through the rules
$${\int}a^{*}{\,}da^{*}{\;}={\;}1{\;};
{\qquad}{\int}a{\,}da{\;}={\;}1{\;};
{\qquad}{\int}da^{*}{\;}={\;}0{\;};
{\qquad}{\int}da{\;}={\;}0{\quad}.\eqno(2.7)$$
\noindent
The inner product of two state vectors is defined to be
$$(f_{1}{\,},{\,}f_{2}){\;}{\,}
={\;}{\,}{\int}\bigl(f_{1}(a^{*})\bigr)^{*}{\;}f_{2}(a^{*}){\;}
e^{-a^{*}a}{\;}da^{*}{\,}da{\quad},\eqno(2.8)$$
\noindent
This inner product is positive-definite.  Indeed, the monomials
$$\psi_{0}{\;}{\,}={\;}{\,}1{\;};
{\qquad}{\quad}\psi_{1}{\;}{\,}={\;}{\,}a^{*}
\eqno(2.9)$$
are an orthogonal pair, each of unit length.  One may interpret 
$\psi_{0}$ as giving a no-particle state and $\psi_{1}$ as giving 
a one-particle state; similarly, the operators ${\bf a^*}$ and 
${\bf a}$ give creation and annihilation operators.  In this way, 
one can replicate the usual structures familiar in a particle 
description of fermions.  But, in the holomorphic representation 
for fermions, the wave functions are encoded through the 
'complex-analytic' form 
${\,}f(a^{*}){\,}={\,}f_{0}{\,}+{\,}f_{1}{\,}a^{*}{\,}$, 
which makes the parallel with bosonic systems much closer;
in the holomorphic representation for bosons [18-20],
wave-functions take the form $f(a^{*}){\,}$, where, 
in the bosonic context, $a^*$ is simply a (commuting) 
complex variable.\par
\smallskip
\indent
This representation can be generalised to fermionic field theory 
[18,19], by considering a collection of Fermi oscillators 
interacting with an external field. In our case (see below), 
we fix half of the spinorial quantities 
${\,}(\phi^{A}(x){\,},{\,}\tilde\phi^{A'}(x)){\,}$ at each point 
$x$ on $\Sigma_I{\,}$, and similarly half at each point on
$\Sigma_F{\,}$, in order to have a well-posed classical boundary-value
problem for the massless Dirac (Weyl) equation.  This amounts to
choosing an initial (and a final) wave-functional, which is a product
of delta-functions in the fermionic variables.  In the simple Fermi
example of Eqs.(2.1-9), one has (for example) 
${\,}\delta(a^{*})=a^*{\,}$.  That is, in our case, the 
wave-functionals specifying the initial and final data are 
(formally) products, over all points $x{\,}$, of suitable functions 
of the type (2.3).  This holomorphic approach for fermions,
of course, exploits the connection between the quantum and the
classical.  Similarly for bosonic fields, where the Bargmann-Fock
holomorphic representation is known as the theory of coherent states
in quantum optics [18-20,28].\par
\medskip
\ni
{\bf 3. Massless spin-${{1}\over{2}}$ fields}
\medskip
\indent
We are concerned with the quantum amplitude to go from initial data
specified on an initial space-like hypersurface $\Sigma_I{\,}$, 
at time $t=0$ (say), measured at spatial infinity, to final data 
on a hypersurface $\Sigma_F{\,}$, at a very late Lorentzian proper 
time $T{\,}$, again measured at spatial infinity.  As described in 
Sec.1, we rotate the time-interval $T$ into the complex: 
$T\rightarrow{\mid}T{\mid}{\,}\exp(-i\theta){\,}$, where 
${\,}0<\theta\leq\pi/2{\,}$.  In this case, the {\it classical} 
boundary-value problem is expected to be strongly elliptic, 
up to gauge, leading to a complex classical solution
$(g_{\mu\nu}{\,},{\,}\phi)$ for Einstein/scalar theory.  From this, 
one can recover the Lorentzian quantum amplitude {\it via}
Feynman's $+i\epsilon$ procedure, by considering the
semi-classical amplitude in the limit 
${\,}\theta\rightarrow 0_{+}{\,}$.  As mentioned above, the
Lorentzian-signature boundary-value problem ${\,}(\theta =0)$ 
is badly posed [2,10,11], even for linear systems of equations 
-- in general, there is no solution, or, if there is a solution, 
it may not be unique.  It is for this reason that calculations in 
quantum field theory (along the lines just given)  must proceed in the
'Euclidean direction', that is, by taking 
${\,}{\rm Im}(T){\,}<{\,}0{\,}$.\par
\ss
\indent
For quantum purposes, then, we should think of the massless
spin-${{1}\over{2}}$ field from the Euclidean viewpoint.  
In the simplest case, one has an unprimed 2-component spinor 
field $\phi^A$ and an independent primed spinor field 
$\tilde\phi^{A'}$ [11] (this reference describes the spinor 
conventions being used here).  In addition to the familiar
Einstein/massless-scalar action of Eq.(3.4) of [5], possibly 
including also a Maxwell contribution as in Eq.(2.1) of [8], 
the (Lorentzian) action ${\,}S$ now contains an extra piece
$$S_{{1}\over{2}}{\;}{\,} 
={\;}{\,}-{\,}{{i}\over{2}}{\;}\int{\,}d^{4}x{\;}{\,}e{\;}{\,}
\tilde\phi^{A'}{\;}e_{AA'}^{~~~~\mu}{\;}\bigl(D_{\mu}\phi^{A}\bigr){\;}
+{\;}{\rm 'h.c.'}+{\;}S_{{1\over 2}B}{\quad}.\eqno(3.1)$$
\ni
Here, as in Sec.6 of [8], ${\,}e_{AA'}^{~~~~\mu}{\,}$ denotes 
the spinor form of the tetrad 
${\,}e_{a}^{~\mu}{\;}(a=0,1,2,3){\,}$, and we define
${\,}e{\,}
={\,}{\rm det}\bigl(e^{a}_{~\mu}\bigr){\,}={\,}(-g)^{{1}\over{2}}{\,}$, 
where ${\,}g{\,}={\,}{\rm det}(g_{\mu\nu}){\,}$.   
The spinor-valued connection one-forms
${\,}\omega^{AB}_{~~~~\mu}{\,}={\,}\omega^{(AB)}_{~~~~~\mu}{\,}$ 
and their tilded counterparts 
${\,}\tilde\omega^{A'B'}_{~~~~~\mu}{\,} 
={\,}\tilde\omega^{(A'B')}_{~~~~~~~\mu}{\,}$ are defined in 
Sec.2.9.2 of [11].  The covariant derivative on spinors is defined,
with the help of the distributive law, {\it via}
$$\eqalignno{D_{\mu}\phi^{A}{\;}{\,}&
={\;}{\,}\partial_{\mu}\phi^A{\;}
+{\,}\omega^{A}_{~~B\mu}{\;}\phi^{B}{\quad},&(3.2)\cr
D_{\mu}\tilde\phi^{A'}{\;}{\,}&
={\;}{\,}\partial_{\mu}\tilde\phi^{A'}{\;}
+{\,}\tilde\omega^{A'}_{~~B'\mu}{\;}\tilde\phi^{B'}{\quad},&(3.3)\cr}$$
\ni
Also, the expression 'h.c.' in Eq.(3.1) denotes an operation in which
$\phi^A$ is replaced by $\tilde\phi^{A'}{\,}$, and 
${\,}\tilde\phi^{A'}{\,}$ by ${\,}\phi^A{\,}$, and in which the order 
of fermionic integrands is reversed.  This is necessary, since 
fermionic quantities such as ${\,}\phi^A{\,}$ or 
${\,}\tilde\phi^{A'}{\,}$ must be taken to be odd elements of a
Grassmann algebra, anti-commuting among each other (obeying, for example,
${\,}\phi^{A}{\,}\tilde\phi^{A'}{\,}={\,}-{\,}\tilde\phi^{A'}\phi^{A}$),
as in Sec.2.  The final element of Eq.(3.1) is a boundary contribution 
$S_{{{1}\over{2}}B}{\,}$, which depends on the fermionic boundary
conditions chosen (see below).\par
\ss
\indent
On varying the independent quantities 
${\,}\phi^A,{\,}\tilde\phi^{A'}{\,}$ in the action (3.1), 
one obtains the Weyl (neutrino) field equations
$$\eqalignno{e_{AA'}^{~~~~\mu}{\;}{\,}D_{\mu}\phi^{A}{\;}{\,} 
&={\;}{\,}0{\quad},&(3.4)\cr
e_{AA'}^{~~~~\mu}{\;}{\,}D_{\mu}\tilde\phi^{A'}{\;}{\,} 
&={\;}{\,}0{\quad}.&(3.5)\cr}$$
\ni
Next, consider possible boundary conditions.  It will again be
simplest and most symmetrical to take local boundary conditions, 
as in the cases of ${\,}s=0{\,}$ [5,6], ${\,}s=1{\,}$ [8] and 
${\,}s=2{\,}$ [9].  By analogy with Sec.6 of [8] and Sec.10 of [9], 
define, at the boundaries $\Sigma_I$ and $\Sigma_F{\,}$:
$$\Phi^{A}_{\ep}(x){\;}{\,}
={\;}{\,}\ep{\;}\sqrt{2}{\;}n^{A}_{~A'}{\;}\tilde\phi^{A'}(x){\;}
+{\;}\phi^{A}(x){\quad},\eqno(3.6)$$
\ni
where ${\,}\ep{\,}={\,}\pm{\;}1{\,}$ and ${\,}n^{AA'}$ is the 
spinor version of the unit time-like future-directed vector 
${\,}n^{\mu}{\,}$, normal to the surface ${\,}\si{\,}$ 
or ${\,}\sf{\,}$.  If one regards the right-hand side of Eq.(3.6) 
as being given by a ${\,}4\times 4$ matrix operator ${\,}P_{\ep}{\;}$, 
acting on a column vector with components
${\,}(\phi^0{\,},\phi^1{\,},{\,}\tilde\phi^{0'},\tilde\phi^{1'}){\,}$, 
one finds that $P_+{\,}$ and $P_-{\,}$ are projection operators, 
obeying
$$P_+{\;}+{\;}P_-{\;}={\;}1{\;},
{\qquad}{\quad}P_+{\,}P_-{\;}{\,}={\;}{\,}P_-{\;}P_+{\;}={\;}0{\;},
{\qquad}{\quad}(P_{\ep})^{2}{\;}={\;}P_{\ep}{\quad}.\eqno(3.7)$$
\ni
The use of such local boundary conditions is suggested by the work of
[24-26].  Given a choice of $\ep{\,}$, take the boundary
contribution ${\,}S_{{{1}\over{2}}B}{\,}$ to the action (3.1) to be
$$S_{{{1}\over{2}}B}{\;}{\,}
={\;}{\,}{{i\epsilon}\over{2}}{\;}\int_{\partial{\cal M}}{\;}d^{3}x{\;}{\,}
h^{{1}\over{2}}{\;}{\,}\Phi^{B}_{\ep}{\;}\phi^{A}{\;}\ep_{AB}
{\quad},\eqno(3.8)$$
\ni
where ${\,}h{\,}={\,}{\rm det}(h_{ij}){\,}$ and $h_{ij}$ gives the 
intrinsic 3-metric of the boundary ${\,}(i,j=1,2,3){\,}$, while 
$\epsilon_{AB}$ is the alternating spinor [22].  One can then verify
that the total action $S_{{1}\over{2}}$ is extremised at a classical
solution, given a specification of ${\,}\Phi^{B}_{\ep}$ at the
boundary.  These are our local boundary conditions.  Note their
similarity to the spin-1 (Maxwell) boundary conditions, involving 
the data 
${\,}\Psi^{AB}_{+}{\,}
=-{\,}2{\,}i{\,}n^{B}_{~B'}{\;}e^{AB'k}{\,}B_k{\,}$ 
on $\si$ and $\sf$ in Eq.(6.18) of [8], which correspond to fixing 
the (spatial) magnetic-field components $B_k{\,}$.  Again, 
for spin-2 (gravitational-wave) perturbations, the boundary
conditions correspond to fixing a projection of the Weyl tensor 
-- see Sec.10 of [9].\par
\ss
\in
At a classical solution, by virtue of the Weyl equations (3.4,5), 
the volume integral vanishes in Eq.(3.1) for the action 
$S_{{1}\over{2}}{\,}$.  Hence, the classical spin-${1}\over{2}$ 
part of the action resides on the boundary and, at lowest 
(quadratic) order in fermions, is given by 
$$\bigl(S_{{1}\over{2}}\bigr)_{\rm class}{\;}{\,} 
={\;}{\,}\bigl(S_{{{1}\over{2}}B}\bigr)_{\rm class}{\quad},\eqno(3.9)$$
\ni
where $S_{{{1}\over{2}}B}$ is given in Eq.(3.8).\par
\ss
\in
For integer-spin perturbations, as in [5,6,8,9], we chose vanishing 
perturbative initial data on the initial space-like hypersurface 
${\,}\si{\,}$, to simplify the exposition.  Correspondingly, 
for massless spin-${{1}\over{2}}$ perturbations, we take
the homogeneous (Dirichlet-type) boundary conditions at $\si$ to be 
$$\Psi^{A}_{-}(x)\Bigl\arrowvert_{\si}{\;}{\,}
={\;}{\,}0{\quad},\eqno(3.10)$$
\ni
(choosing ${\,}\ep{\,}={\,}-{\,}1$ here).  Equivalently,
$$\phi^{A}{\;}{\,} 
={\;}{\,}{\sqrt 2}{\;}{\,}n^{A}_{~A'}{\;}\tilde\phi^{A'}\eqno(3.11)$$
\ni
on $\si{\,}$.  Hence, there is no contribution to the classical 
action from the initial surface.  On the final hypersurface $\sf{\,}$, 
we specify (non-zero) data $\Psi^{A}_{+}(x){\,}$, corresponding 
to the presence of a remnant flux of neutrinos outgoing at large 
radius and late time:
$$\Psi^{A}_{+}(x)\Bigl\arrowvert_{\sf}{\;}{\,}
={\;}{\,}f^{A}(r,\Omega){\quad}.\eqno(3.12)$$ 
\ni
In this case, unless one has found a complete solution of the field
equations, $\Psi^{A}_{-}{\,}$ is (as yet) unknown on $\sf{\,}$, 
while $\Psi^{A}_{+}{\,}$ is unknown on $\si{\,}$.\par
\ss
\in
Following the discussion in Sec.2, the particular choice
(3.10), with zero right-hand side, would be interpreted as containing
no particles on $\si{\,}$.  But the choice (3.12) of final data, for
$f^{A}\,{\neq}\,0{\,}$, specifies that a certain amount of fermions were
produced, to arrive at $\sf{\,}$.  Of course, one can specify the
right-hand sides of Eqs.(3.10,12) at will (subject to fall-off at
infinity) to investigate the effects of allowing incoming fermions on
$\si$ and asking how they contribute to the final fermionic output at
$\sf{\,}$.  Because the classical fermionic action in Eq.(4.35) below
is 'bosonic' -- that is, it is analogous to the particular wave
function $f(a^{*},a)=f_{11}\,aa^{*}$ in Eq.(2.2) -- fermions can only be
produced or destroyed in pairs.  For example, from Eq.(4.35) it is
impossible to have precisely one neutrino in the far future if none
were present in the far past.\par
\ss
\in
In our case, fermionic boundary data are to be specified on a 
pair of asymptotically-flat space-like hypersurfaces $\si$ 
and $\sf{\,}$, together with the boundary data, which include the
asymptotic proper-time interval $T{\,}$, as measured at spatial
infinity.  It turns out [11] that one should then specify
$\Psi^{A}_{-}$ on one surface and $\Psi^{A}_{+}$ on the other, 
to have a well-posed classical problem.  In the 'cosmological' 
case, where, for example, a three-sphere ${\Bbb S}^3$ bounds a 
four-ball ${\Bbb B}^4{\,}$, different considerations may 
apply [11,29-31].\par 
\medskip
\ni
{\bf 4. Summary of the Vaidya approximation}
\ms
\in
By analogy with the ${\,}s=0$ case in [1-7], we now study weak 
$s={{1}\over{2}}$ massless (neutrino) fields in a 
spherically-symmetric 'background' Lorentzian-signature geometry
$$ds^{2}{\;}{\,}
={\;}{\,}-{\,}e^{b(t,r)}{\,}dt^{2}{\,}+{\,}e^{a(t,r)}{\,}dr^{2}{\,}
+{\,}r^{2}{\,}\Bigl(d\theta^{2}{\,}
+{\,}\sin^{2}\theta{\;}d\phi^{2}\Bigr){\quad},\eqno(4.1)$$
\noindent  
along the lines described in [1,2].  We are principally interested 
in quantum emission of a roughly stochastic type, following 
gravitational collapse to a black hole, since generically this leads 
to the most probable final configurations.  At late times, 
as described further in [1-9], the configuration of all fields 
(not just the 'background' spherically-symmetric coupled solution 
of the Einstein/massless-scalar field equations, but also the
averaged second-variation energy-momentum tensor $T_{\mu\nu}$ 
generated by the anisotropic perturbative fields) will approximate
closely a nearly-spherically-symmetric stream of radially-outgoing
null or massless radiation, together with the gravitational field
which this generates.  This is treated in detail in [3,4].
Here, we summarise briefly the necessary properties of such an approximately
Vaidya metric [7,27].  In the companion paper [4] on spins 1 and 2, a
full description of the aspects of the Vaidya solution relevant to the
present spin-1/2 case is given.  In [7], moreover, a detailed
treatment of the Vaidya metric, as needed for the general question of
black-hole evaporation, can be found.  In addition to treatment of the
Vaidya-like 
geometry in which the particles or fields propagate, [7] also includes
a thorough description of the energy-momentum source for the
Vaidya-like geometry, produced by a large number of null 
(radiating) particles. \par  
\ss
\in
As in [1-9], we write $e^{-a(t,r)}$ in terms of the 
'mass function' $m(t,r)$ as
$$e^{-a(t,r)}{\;}{\,}
={\;}{\,}1{\,}-{{2m(t,r)}\over r}{\quad}.\eqno(4.2)$$ 
\ni
A Vaidya metric [7,27] corresponds to a spherically-symmetric 
null-fluid energy-momentum source, of the form
$$T_{\mu\nu}{\;}{\,}={\;}{\,}f(t,r){\;}k_{\mu}{\,}k_{\nu}{\quad}.
\eqno(4.3)$$
Here, ${\,}f(t,r){\,}$ is some spherically-symmetric function, 
and the vector field ${\,}k^{\mu}{\,}$ is null ($k^{\mu}k_{\mu}=0$) 
and points radially outward to the future (in our case).  As is
described much more fully in the companion paper [4], the
corresponding Vaidya solution of the Einstein field equations with the
energy-momentum source (4.3) can be put in the diagonal form (4.1), as  
$$ds^{2}{\;}{\,} 
={\;}-{\;}\biggl({\dot m \over m'}\biggr)^{2}{\;}
{\biggl(1-{{2m(t,r)}\over r}\biggr)}^{-1}{\;}dt^{2}{\;}
+{\;}\biggl(1-{{2m(t,r)}\over r}\biggr)^{-1}{\;}dr^{2}{\,}
+{\,}r^{2}{\;}d\Omega^{2}{\quad}.\eqno(4.4)$$ 
\ni
Here, ${\,}m=m(t,r){\,}$, and ${\,}{\dot m},{\;}m'$ denote 
${\,}{\partial m}/{\partial t}{\,},{\;}{\partial m}/{\partial r}{\,}$, 
respectively.  Thus, 
$$e^{b(t,r)}{\;}{\,}
={\;}{\,}\biggl({\dot m\over m'}\biggr)^{2}{\;}
\biggl(1-{2m\over r}\biggr)^{-1}{\quad}.\eqno(4.5)$$
Moreover, as seen in [4,7], the Einstein field equations imply that  
$$m'{\;}\biggl(1-{{2m}\over{r}}\biggr){\;}{\,}={\;}{\,}f(m){\quad},
\eqno(4.6)$$
\ni
where ${\,}f(m){\;}{\ge}{\;}0{\,}$ is an arbitrary (smooth) function.
Physically, in our example, the function $f(m)$ in Eq.(4.6) depends on
the specific physical (particle) model.  This metric describes the 
space-time region in which radiation is streaming radially outwards, 
due to the black-hole evaporation.  Correspondingly, the 
'mass function' $m(t,r)$ varies extremely slowly with time, and 
effectively one has a Schwarzschild metric with a very slowly-varying 
mass, in which $f(m)$ can be regarded as nearly constant.  
As described in the preceding paper [4], the geometry will gradually 
deviate from the Vaidya form as one moves to the past into the 
strongly dynamical collapse region.  Provided that the complexified 
classical boundary-value problem is well-posed, the complexified 
classical solution $(g_{\mu\nu},\phi)$ will be regular at every point, 
including the spatial origin $r=0{\,}$.\par
\smallskip
\indent
As further described in [4,7], a coordinate transformation can be
found, which leads to the usual form 
$$ds^{2}{\;}{\,}
={\;}{\,}-\Biggl(1{\,}-{\,}{{2m(u)}\over{r}}\Biggr){\,}du^{2}{\,}
-{\,}2{\,}du{\,}dr{\,}
+{\,}r^{2}{\,}\bigl(d{\theta}^{2}{\,}
+{\,}{\sin}^{2}\theta{\,}d{\phi}^{2}\bigr){\quad},\eqno(4.7)$$
\noindent
of the Vaidya metric [7,27], where $m= m(u)$ is now seen to be a
function of a single null variable $u$.  The diagonal form (4.4)
will be used below, in studying the evolution of linearised 
perturbation fields on the slowly-varying sequence of 
Schwarzschild geometries which is described  closely here by a Vaidya
metric.  The 'null' form (4.7) is instead useful in understanding the
large-scale structure of the Vaidya space-time.\par
\ss
\in
From the Vaidya metric in the form (4.7), one can see that the 
apparent singularity in the metric (4.4) at ${\,}r=2m(u){\,}$ is only 
a coordinate singularity [27].  Further, the surface $\{r=2m(u)\}$ 
is space-like, lying to the past of the region $\{r>2m(u)\}$.  
In fact, the geometry in the region $\{r<2m(u)\}$ will gradually 
deviate from the Vaidya form, as one moves to the past by (say) 
reducing $u$ while holding $r$ fixed, since one reaches the region 
of strong-field gravitational collapse.  This region can still be 
described by the diagonal metric (4.1), with scalar field 
${\,}\phi =\phi(t,r){\,}$, but the full field equations enforce 
a more complicated coupled solution.  Again, provided that the 
complexified boundary-value problem of Sec.1 remains well-posed for 
${\,}0<\theta\leq{\pi}/2{\,}$, the full, complexified Einstein/scalar
classical solution studied here will be regular at the spatial origin 
${\,}r=0{\,}$.\par
\ms
\ni
{\bf 5. Separation of neutrino equations}
\ms
\in
Separation of variables for the neutrino equation in a Kerr background
space-time was carried out by Teukolsky [32] and by Unruh [33], using
the Newman-Penrose formalism [34].  As in [5,6] for $s=0{\,}$, 
[8] for $s=1$ and [9] for $s=2{\,}$, we are principally concerned 
here with the wave-like (Lorentzian) evolution of neutrino fields 
in the Vaidya-like metric [7,27], which describes the extremely
slow evolution of the black hole, as it gradually loses mass through
nearly-isotropic radiation.  For most radiation frequencies of interest,
the time rate of change of the background geometry will be so slow as
to be negligible.  Thus, the evolution of the wave will be adiabatic.\par
\ss
\in
Following [32,33], we decompose the neutrino field 
$\bigl(\phi^{A},\tilde\phi^{A'}\bigr)$ with respect to the normalised 
spinor dyad $(\iota^{A}{\,},{\,}o^{A})$ which corresponds 
to the Kinnersley null tetrad [35] in the Schwarzschild metric.  This
pair $(o^{A},{\,}\iota^{A})$ gives a basis for the 2-complex-dimensional
vector space of spinors $\omega^A$ at each point, and is normalised 
according to
$o_{A}{\;}\iota^{A}{\;}{\,} 
={\;}{\,}1{\;}{\,} 
={\;}{\,}-{\,}\iota_{A}{\;}o^{A}{\quad}$.  
Knowledge of the Kinnersley null tetrad [34,35]
${\,}\ell^{\mu}{\,},{\,}n^{\mu}{\,},{\,}m^{\mu}{\,},{\,}\bar m^{\mu}$ 
of vectors at a point is equivalent to knowledge of the corresponding 
normalised spinor dyad $(o^{A},\;\iota^{A})$, through the relations
${\,}l^{\mu}{\;}\leftrightarrow{\;}o^{A}{\;}o^{A'}, 
{\,}n^{\mu}{\;}\leftrightarrow{\;}\iota^{A}{\;}\iota^{A'},
{\,}m^{\mu}{\;}\leftrightarrow{\;}o^{A}{\;}\iota^{A'},
{\,}{\bar m}^{\mu}{\;}\leftrightarrow{\;}\iota^{A}{\;}o^{A'}$.\par
\ss
\in
Writing 
${\,}\phi^{A}{\,}
={\,}\chi^{(0)}{\,}o^{A}{\,}+{\,}\chi^{(1)}{\,}\iota^{A}{\,}$, 
one obtains the dyad version of the field equation
$e_{AA'}^{~~~~\mu}{\;}D_{\mu}\phi^{A}{\,}={\,}0{\,}$:  
$$\eqalignno{{{1}\over{r\sqrt2}}{\,}
\biggl(\partial_{\theta}{\,}+{\,}{{i}\over{\sin\theta}}{\;}\partial_{\phi}{\,} 
+{{1}\over{2}}{\,}\cot\theta\biggr){\,}\chi^{(0)}{\,}+{\,}{{1}\over{2}}{\,}
\biggl(\partial_{t}{\,}-e^{-a}{\;}\partial_{r}{\,}
-{\,}{{(r-m)}\over{r^{2}}}\biggr)\chi^{(1)}{\;}
&={\;}0{\;},&(5.1)\cr
\biggl(e^{a}{\,}\partial_{t}{\,}+{\,}\partial_{r}{\,}
+{\,}{{1}\over{r}}\biggr){\,}\chi^{(0)}{\,}
+{\,}{{1}\over{r\sqrt 2}}{\,}\biggl(\partial_{\theta}{\,}
+{\,}{{i}\over{\sin\theta}}{\;}\partial_{\phi}{\,}
+{\,}{{1}\over{2}}{\,}\cot\theta\biggr)\chi^{(1)}{\;}
&={\;}0{\;}.&(5.2)\cr}$$
\ss
\indent
As is standard, one can separate variables in Eqs.(5.1,2).  We write
$$\eqalignno{\chi^{(0)}(t,r,\theta,\phi){\;}{\,}
&={\;}{\,}{1\over r}{\;}\sum^{\infty}_{\ell={{1}\over{2}}}{\;}
\sum^{\ell}_{m=-1}{\;}S_{-{{1}\over{2}}\ell m}(\theta){\;}{\,}
e^{im\phi}{\;}{\,}R_{-{{1}\over{2}}\ell m}(t,r){\quad},&(5.3)\cr
\chi^{(1)}(t,r,\theta,\phi){\;}{\,}
&={\;}{\,}\sum^{\infty}_{\ell={{1}\over{2}}}{\;}\sum^{\ell}_{m=-1}{\;}
S_{+{{1}\over{2}}\ell m}(\theta){\;}{\,}e^{im\phi}{\;}{\,}
R_{+{{1}\over{2}}\ell m}(t,r){\quad},&(5.4)\cr}$$
\ni
From Eqs.(5.3,4), ${\,}S_{\pm{{1}\over{2}}\ell m}(\theta,\phi)$ and   
${\,}R_{\pm{{1}\over{2}}\ell m}(\theta,\phi)$ satisfy:
$$\eqalignno{{{1}\over{2}}{\;}r^{2}{\,}
\biggl(\partial_{t}{\,}-{\,}e^{-a}{\;}\partial_{r}{\,}
-{\,}{{(r-M)}\over{r^{2}}}\biggr){\;}R_{+{{1}\over{2}}\ell m}{\;}{\,}
&={\;}{\,}-{\,}\lambda_{1}{\;}R_{-{{1}\over{2}}\ell m}{\quad},&(5.5)\cr
{{1}\over{\sqrt 2}}{\,}\biggl(\partial_{\theta}{\,} 
-{\,}{{m}\over{\sin\theta}}{\,}+{\,}{{1}\over{2}}{\,}\cot\theta\biggr){\;}
S_{-{{1}\over{2}}\ell m}{\;}{\,}
&={\;}{\,}\lambda_{1}{\;}S_{+{{1}\over{2}}\ell m}{\quad},&(5.6)\cr
\bigl(e^{a}{\,}\partial_{t}{\,}+{\,}\partial_{r}\bigr){\,}
R_{-{{1}\over{2}}\ell m}{\;}{\,}
&={\;}{\,}-{\,}\lambda_{2}{\;}R_{+{{1}\over{2}}\ell m}{\quad},&(5.7)\cr
{{1}\over{\sqrt 2}}{\,}\biggl(\partial_{\theta}{\,} 
+{\,}{{m}\over{\sin\theta}}{\,}+{\,}{{1}\over{2}}{\,}\cot\theta\biggr){\,}
S_{+{{1}\over{2}}\ell m}{\;}{\,}
&={\;}{\,}-{\,}\lambda_{2}{\;}S_{-{{1}\over{2}}\ell m}{\quad},
&(5.8)\cr}$$
\ni
where $\lam_{1}$ and $\lam_{2}$ are separation constants.  
Regularity of the angular functions ${\,}S_{\pm{{1}\over{2}}\ell m}$ 
at $\theta =0{\,}$ and ${\,}\theta =\pi$ implies that
$\lam_{1}=\lam_{2}{\,},{\;}=\lam_{\ell}{\,}$, say.  That is, 
if we set ${\,}\theta'=(\pi -\theta)$ in Eq.(5.6,8), we find that   
${\,}S_{\pm{{1}\over{2}}\ell m}(\theta){\;}
\propto{\;}S_{\mp{{1}\over{2}}\ell m}(\pi -\theta){\,}$.  
Up to normalisation, the functions 
${\,}S_{\pm{{1}\over{2}}\ell m}{\;}{\,}e^{im\phi}$ are just spin-weighted
spherical harmonics ${\,}_{s}Y_{\ell m}(\Omega)$ for 
${\,}s=\pm{{1}\over{2}}{\;}$ [36,37].  These functions have the properties
$$\eqalignno{_{s}Y^{*}_{\ell m}{\;}{\,}
&={\;}{\,}(-1)^{m+s}{\;}{\,}_{-s}Y_{\ell,-m}{\quad},&(5.9)\cr
_{s}Y_{\ell m}(\pi -\theta{\,},{\,}\pi +\phi){\;}{\,} 
&={\;}{\,}_{s}Y_{\ell m}(\theta,\phi){\quad}, &(5.10)\cr
\int{\,}d\Omega{\;}{\,}_{s}Y_{\ell m}(\Omega)
{\;}{\,}_{s}Y^{*}_{\ell' m'}(\Omega){\;}{\,}
&={\;}{\,}\delta_{\ell \ell'}{\;}\delta_{mm'}{\quad}.&(5.11)\cr}$$
\ni
In general, the $S_{s\ell m}{\,}$, which are regular on $[0,\pi]{\,}$,
satisfy:
$$\eqalign{{{1}\over{\sin\theta}}{\;}{{d}\over{d\theta}}
&\biggl(\sin\theta{\;}{{dS_{s\ell m}}\over{d\theta}}\biggr)\cr
&+{\,}\biggl({{-m^{2}}\over{\sin^{2}\theta}}{\,} 
-{\,}{{2ms{\,}\cot\theta}\over{\sin\theta}}{\,}
-{\,}s^{2}\cot^{2}\theta{\,}+{\,}s{\,}+{\,}\nu\biggr){\,}S_{s\ell m}{\;}{\,} 
={\;}{\,}0{\quad},\cr}\eqno(5.12)$$
\ni
where ${\,}\nu{\,}={\,}(\ell -s)(\ell +s+1){\,}$.  For our case 
${\,}{\mid}s{\mid}={{1}\over{2}}{\,}$, one has 
${\,}\nu{\,}
={\,}2{\,}(\lam_{\ell})^{2}{\,}
={\,}(\ell +{1\over 2})^{2}{\,}$.\par
\ss
\in
Finally, we make a further change of variables to
$$\eqalignno{P_{+{{1}\over{2}}\ell m}(t,r){\;}{\,}
&={\;}{\,}{{r}\over{\sqrt 2}}{\;}e^{-a/2}{\;}{\,}
R_{+{{1}\over{2}}{\ell m}}(t,r){\quad},&(5.13)\cr
P_{-{{1}\over{2}}{\ell m}}(t,r){\;}{\,}
&={\;}{\,}R_{-{{1}\over{2}}{\ell m}}(t,r){\quad},&(5.14)\cr}$$ 
\ni
and then to
$$\eqalignno{P_{-{{1}\over{2}}{\ell m}}(t,r){\;}{\,}
&={\;}{\,}{{1}\over{2{\,}\sqrt{\ell+{1\over 2}}}}{\;}
\Bigl(\xi_{{{1}\over{2}}\ell m+}(t,r){\;} 
-{\;}\xi_{{{1}\over{2}}{\ell m-}}(t,r)\Bigr){\quad},&(5.15)\cr
P_{+{{1}\over{2}}{\ell m}}(t,r){\;}{\,}
&={\;}{\,}{{1}\over{2{\,}\sqrt{\ell+{{1}\over{2}}}}}{\;}
\Bigl(\xi_{{{1}\over{2}}\ell m+}(t,r){\;} 
+{\;}\xi_{{{1}\over{2}}{\ell m-}}(t,r)\Bigr){\quad}.&(5.16)\cr}$$
\ni
We obtain the more symmetrical set of equations
$$\eqalignno{\biggl(e^{-a}{\;}\partial_{r}{\,}
+{\,}{{\lam_{\ell}{\,}\sqrt2}\over{r}}{\;}e^{-a/2}\biggr){\,}
\xi_{{{1}\over{2}}{\ell m-}}{\;}{\,} 
&={\;}{\,}\partial_{t}\xi_{{{1}\over{2}}\ell m+}{\quad},&(5.17)\cr
\biggl(e^{-a}{\;}\partial_{r}{\,}
-{\,}{{\lam_{\ell}{\,}\sqrt2}\over{r}}{\;}e^{-a/2}\biggr){\,}
\xi_{{{1}\over{2}}{\ell m+}}{\;}{\,} 
&={\;}{\,}\partial_{t}\xi_{{{1}\over{2}}\ell m-}{\quad}.&(5.18)\cr}$$
\ni
These yield the second-order decoupled  $(t,r)$ form of the neutrino 
wave equation
$$e^{-a}{\;}\partial_{r}\bigl(e^{-a}{\;}
\partial_{r}{\,}\xi_{{{1}\over{2}}\ell m\pm}\bigr){\;} 
-{\;}\bigl(\partial_{t}\bigr)^{2}{\,}\xi_{{{1}\over{2}}\ell m\pm}{\,}
-{\,}V_{{{1}\over{2}}\ell\pm}{\;}{\,}\xi_{{{1}\over{2}}\ell m\pm}{\;}{\,}
={\;}{\,}0{\quad},\eqno(5.19)$$
\ni
where the real potential(s) ${\,}V_{{{1}\over{2}}\ell\pm}{\,}$ 
are given by
$$V_{{{1}\over{2}}\ell\pm}{\;}{\,}
={\;}{\,}e^{-a}{\,}\biggl({{2{\,}(\lam_{\ell})^{2}}\over{r^{2}}}{\;}\pm{\;}
\lam_{\ell}{\;}\sqrt{2}{\;}{\,}\partial_{r}
\biggl({{e^{-a/2}}\over{r}}\biggr)\biggr){\quad}.\eqno(5.20)$$
\ss
\in
Thus, we have expressed the perturbative neutrino field in terms of
'odd' and 'even', or 'magnetic' and 'electric' components 
${\,}\xi_{{{1}\over{2}}\ell m\pm}{\,}$, just as we did with the 
integer-spin perturbations.  One can now use the Kinnersley tetrad
$$\eqalign{\ell^{\mu}{\;}{\,}
&={\;}{\,}\biggl(1/\Bigl(1-{{2m}\over{r}}\Bigr){\,},
1{\,},0{\,},0\biggr){\quad},\cr 
n^{\mu}{\;}{\,}
&={\;}{\,}\biggl(1{\,},-{\,}\Bigl(1-{{2m}\over{r}}\Bigr){\,},
0{\,},0\biggr)/2{\quad},\cr
m^{\mu}{\;}{\,}
&={\;}{\,}\biggl(0{\,},0{\,},1{\,},
\Bigl(i/\sin\theta\Bigr)\biggr)/r\sqrt2{\quad},\cr}\eqno(5.21)$$   
\ni
in the Schwarzschild metrics together with the boundary condition 
(3.10), to write the classical action (3.8,9) as
$$\eqalign{(S_{{{1}\over{2}}B})_{\rm class}{\;}{\,}
&={\;}{\,}{{i}\over{2}}{\,}\int{\,}d\Omega{\,}\int^{R\infty}_{0}{\,}dr{\;}{\,}
r^{2}{\;}e^{a/2}{\;}\bigl(\Psi^{B}_{+}{\;}o_{B}{\;}\chi^{(0)}{\,}
+{\,}\Psi^{B}_{+}{\;}\iota_{B}{\;}\chi^{(1)}\bigr)\Bigl\arrowvert_{\sf}\cr
&={\;}{\,}{{i}\over{\sqrt 2}}{\;}\sum^{\infty}_{\ell ={{1}\over{2}}}{\;}
\sum^{\ell}_{m=-1}{\;}
{{\bigl(\ell-{{1}\over{2}}\bigr)!}\over{\bigl(\ell +{{1}\over{2}}\bigr)!}}{\;}
\int^{R\infty}_{0}{\,}dr{\;}{\,}e^{a}{\;}{\,} 
\tilde\xi_{{{1}\over{2}}\ell m-}{\;}{\,}\xi_{{{1}\over{2}}\ell m-}
\Bigl\arrowvert_{T}{\quad}.\cr}\eqno(5.22)$$
\ni
Here we have also used Eqs.(5.9,11,15,16).\par
\ss
\in
The boundary condition (3.10) translates into
$$\xi_{{{1}\over{2}}\ell m+}\Bigl\arrowvert_{t=0}{\;}{\,}
={\;}{\,}0{\quad}.\eqno(5.23)$$
\ni
From Eq.(5.21), one finds
$$\Bigl(\partial_{t}\xi_{{{1}\over{2}}\ell m-}\Bigr)
\Bigl\arrowvert_{t=0}{\;}{\,}
={\;}{\,}0{\quad}.\eqno(5.24)$$
\ni
In addition, one has
$$\eqalignno{\Psi^{A}_{+}{\;}o_{A}{\;}{\,}
&={\;}{\,}\chi^{(1)}{\;}+{\;}\sqrt2{\;}e^{a/2}{\;}\tilde\chi^{(0')}{\;}{\,} 
={\;}{\,}{{\sqrt2{\;}e^{a/2}}\over{r}}{\;}{\,}\sum_{\ell m}{\;}
{\,}_{+{{1}\over{2}}}Y_{\ell m}(\Omega){\;}{\,}
{{\xi_{{{1}\over{2}}\ell m-}(t,r)}\over{\sqrt{\ell +{{1}\over{2}}}}}
{\quad},&(5.25)\cr
\Psi^{A}_{+}{\;}\iota_{A}{\;}{\,}
&={\;}{\,}{{e^{-a/2}}\over{\sqrt2}}{\;}{\,}\tilde\chi^{(1')}{\;}
-{\;}\chi^{(0)}{\;}{\,}
={\;}{\,}{{1}\over{r}}{\;}\sum_{\ell m}{\;}
{\,}_{-{{1}\over{2}}}Y_{\ell m}(\Omega){\;}{\,}
{{\xi_{{{1}\over{2}}\ell m-}(t,r)}\over{\sqrt{\ell +{{1}\over{2}}}}}
{\quad}.&(5.26)\cr}$$
\ni
Accordingly, we set
$$\xi_{{{1}\over{2}}\ell m-}(t,r){\;}{\,}
={\;}{\,}\int^{\infty}_{-\infty}{\;}dk{\;}{\,}
c_{{{1}\over{2}}k\ell m-}{\;}{\,}
\xi_{{{1}\over{2}}k\ell -}(r){\;}{\,}{{\cos(kt)}\over{\cos(kT)}}{\quad},
\eqno(5.27)$$
\ni
where the ${\,}\{c_{{{1}\over{2}}k\ell m-}\}$ are certain coefficients 
taking (odd) values in a Grassmann algebra, and the
${\,}\{\xi_{{{1}\over{2}}k\ell-}(r)\}$ are radial functions.  
As above, we have assumed that, in a neighbourhood of $\sf{\,}$, 
an adiabatic approximation is valid, such that the frequencies 
are large when compared with the inverse time-scale for the very slow 
change in the background metric.\par
\ss
\in
From Eqs.(5.17-19), regularity near the centre of symmetry 
${\,}\{r=0\}$ demands that
$$\xi_{{{1}\over{2}}k\ell\ep}(r){\quad}
\sim{\quad}{\mu}_{k\ell\ep}{\;}{\,}r{\;}{\,}
j_{\ell -{{1}\over{2}}\ep}(kr){\quad},\eqno(5.28)$$
\ni
as ${\,}r\rightarrow 0{\,}$, where 
${\,}j_{\ell -{{1}\over{2}}\ep}(kr)$ denotes a spherical Bessel 
function [38], and ${\,}{\mu}_{k\ell\ep}$ is a normalisation 
constant.  Hence, the ${\,}\{\xi_{{{1}\over{2}}k\ell\ep}(r)\}$
are real functions.\par
\ss
\in
Noting the $(t,r)$ neutrino field equation (5.19), we set
$$\xi_{{{1}\over{2}}k\ell -}(r){\;}{\,}
={\;}{\,}\Bigl(\bigl(z_{{{1}\over{2}}k\ell -}\bigr){\,}e^{ikr^{*}}{\,}
+{\;}\bigl(z^{*}_{{{1}\over{2}}k\ell -}\bigr){\,}e^{-ikr^{*}}\Bigr){\;}
f_{{{1}\over{2}}k\ell -}(r){\quad},\eqno(5.29)$$
\ni
where the ${\,}z_{{{1}\over{2}}k\ell -}$ are complex coefficients, and
we take the boundary condition at infinity to be
${\,}f_{{{1}\over{2}}k\ell -}(r)\rightarrow 1{\,}$ as 
${\,}r\rightarrow\infty{\,}$, so that 
${\,}\xi_{{{1}\over{2}}k\ell-}(r){\;}{\,}
\sim{\;}{\,}\Bigl(\bigl(z_{{{1}\over{2}}k\ell -}\bigr){\,}e^{ikr^{*}}{\,}
+{\;}\bigl(z^{*}_{{{1}\over{2}}k\ell -}\bigr){\,}e^{-ikr^{*}}\Bigr)$ 
as ${\,}r\rightarrow\infty{\,}$.  Here, ${\,}r^{*}$ is a generalised 
Regge-Wheeler coordinate [39,40], such that
$${{dr^{*}}\over{dr}}{\;}{\,}={\;}{\,}e^{(a-b)/2}{\quad},\eqno(5.30)$$
\ni
and one also has
${\,}f_{{{1}\over{2}},-k\ell -}(r){\,}
={\,}f_{{{1}\over{2}}k\ell -}(r){\,}$.  
The coefficients ${\,}z_{{{1}\over{2}}k\ell -}{\,}$ satisfy 
${\;}z^{*}_{{{1}\over{2}}k\ell -}{\,}
={\,}z_{{{1}\over{2}},-k\ell -}{\;}{\,}$.  
The normalisation of the radial functions is given by
$$\int^{R\infty}_{0}{\,}dr{\;}{\,}e^{a}{\;}{\,}
\tilde\xi_{{{1}\over{2}}k\ell -}(r){\;}{\,}
\xi_{{{1}\over{2}}k'\ell -}(r){\;}{\,}
={\;}{\,}2\pi{\;}\tilde z_{{{1}\over{2}}k\ell -}{\;}{\,}
z_{{{1}\over{2}}k\ell -}{\;}{\,}
\Bigl(\delta(k,k'){\,}+{\,}\delta(k,-k')\Bigr){\quad}.\eqno(5.31)$$   
\ni
Hence, the classical spin-${{1}\over{2}}$ action (5.22) can be 
written in the form
$$\eqalign{&(S_{{{1}\over{2}}B})_{\rm class}{\;}{\,}
={\;}{\,}-i{\,}\pi{\,}{\sqrt 2}{\;}{\,}
\sum^{\infty}_{\ell={{1}\over{2}}}{\,} 
\sum^{\ell}_{m=-\ell}{\;}
{{(\ell-{{1}\over{2}})!}\over{(\ell +{{1}\over{2}})!}}{\quad}\times\cr
&{\quad}\times{\;}\int^{\infty}_{0}{\,}d\omega{\;}{\,}
{\tilde z}_{{{1}\over{2}}\omega\ell -}
{\;}{\,}z_{{{1}\over{2}}\omega\ell -}{\;}
\biggl(\Bigl({\tilde c}_{{{1}\over{2}}\omega\ell m-}\Bigr)
+\Bigl({\tilde c}_{{{1}\over{2}}-\omega\ell m-}\Bigr)\biggr)
\biggl(\Bigl(c_{{{1}\over{2}}\omega\ell m-}\Bigr)
+\Bigl(c_{{{1}\over{2}}-\omega\ell m-}\Bigr)\biggr){\;}.\cr}
\eqno(5.32)$$
\ss
\in
This is the desired form of the spin-${{1}\over{2}}$ classical action,
providing the nearest correspondence with the bosonic expressions
Eq.(3.8) of [6], Eq.(5.19) of [10] and Eq.(9.7) of [9] for spins 
$s=0{\,},1{\,}$ and $2{\,}$, respectively.  When one compares this 
with the massless-scalar (second-variation) action 
$S_{\rm class}$ in Eq.(3.8) of [6], the $s=0$ expression involves 
an integral $\int^{\infty}_{-\infty}{\,}dk{\;}k{\,}(~~){\,}$, 
whereas for $s={{1}\over{2}}$ we have 
$\int^{\infty}_{0}{\,}d\omega{\;}(~~){\,}$, with no power of 
$\omega{\,}$.  The $s=0$ integrand is schematically of the
form ${\tilde z}{\,}z{\,}{\tilde a}{\,}a{\,}$, whereas for 
$s={{1}\over{2}}$ one has ${\tilde z}{\,}z{\,}{\tilde c}{\,}c{\,}$, 
where the tildes refer to the primed spinor field $\tilde\phi^{A'}$, 
and untilded quantities refer to the unprimed field $\phi^A{\,}$; 
this structure is inevitable in massless fermionic theories.  
Finally, the $s=0$ integrand includes a factor $\cot(kT)$, which was 
crucial in determining the form of the $s=0$ quantum amplitude in [6], 
{\it via} rotation of $T$ into the complex.  In the $s={{1}\over{2}}$ 
case, this factor of $\cot(kT)$ is absent, and the calculation of the 
$s={{1}\over{2}}$ quantum amplitude is correspondingly simpler.\par
\medskip
\ni
{\bf 6. Conclusion}
\medskip
\in
We have seen in this paper how the calculations in [5-9], treating 
quantum amplitudes for the emission of bosonic particles $(s=0,1,2)$, 
in a nearly-spherically-symmetric Einstein/massless-scalar collapse 
to a black hole, can be extended to the fermionic case of a massless
spin-${{1}\over{2}}$ neutrino field.  Following the treatment 
in [1-3,5-7] of the $s=0$ massless-scalar case, the classical action 
functional for $s={1\over2}$ data on the final late-time space-like 
hypersurface $\sf{\,}$ has been calculated.  One again takes the 
initial Einstein/scalar data on $\si$ to be exactly spherically 
symmetric, describing (say) diffuse, slowly-moving initial data 
for gravitational collapse to a black hole.  We assume that the 
(real) Lorentzian proper-time separation $T$ between ${\,}\si$ 
and ${\,}\sf{\,}$, as measured at spatial infinity, 
is sufficiently large that ${\,}\sf$ intersects all the radiation 
emitted.  Moreover, the final data consist of a section of a 
spherically-symmetric approximately Vaidya geometry, together with 
weak spin-$1\over 2$ data.  We rotate
$T\rta{\mid}T{\mid}\exp(-i\theta)$ into the lower-half complex-$T$ 
plane $(0<\theta\leq\pi/2)$, after Feynman.  We expect that the 
classical boundary-value problem will then be well-posed.
The classical action has here been calculated as a functional 
of the (suitably chosen) boundary data, leading to the 
quantum amplitude.\par
\ss
\in
Perhaps the main differences between the bosonic calculations 
and the present fermionic calculation are that, classically, 
the spin-${{1}\over{2}}$ field is described by odd 
(anti-commuting) Grassmann quantities, and, of course, that the 
field equations are of first order, unlike the commuting and
second-order nature of the bosonic systems. The first-order
property for spin-$1\over 2$ is reflected in our choice of 
local boundary conditions, 'inherited' from local 
supersymmetry [24-26,41,42].  This 'ancestral' appearance 
of supergravity is reflected in the form of the classical action 
functional of the corresponding 'local' boundary data, as well as 
in the resulting quantum amplitude.  An analogous 
(but more complicated) treatment of amplitudes for the 
spin-${{3}\over{2}}$ field, crucial for local supersymmetry 
(supergravity) will soon be completed [43].\par
\ss
\in
As remarked in [2], the present approach leads to a calculation of
quantum amplitudes, appropriate for studying the black-hole radiation
which follows gravitational collapse.  This formulation must be
different from the familiar one (which is normally carried out by
considering Bogoliubov transformations), since it yields quantum
amplitudes relating to the final state, and not just the usual
probabilities for outcomes at a late time and large radius.  It is not
that many of the attributes of the radiation, whether described in the
present way or in the usual way, can be very different in the two
descriptions.  It is rather that the present description allows one to
ask more questions of the radiation (and to receive a response).\par 
\parindent = 1 pt
\medskip
{\bf Acknowledgements}
\medskip
\noindent
We are grateful to the referee for questions and suggestions which led
to substantial improvements in the paper.
\medskip
\noindent
{\bf References}
\medskip
\indent
[1] A.N.St.J.Farley and P.D.D'Eath, 
Phys Lett. B {\bf 601}, 184 (2004).\par           
\indent
[2] A.N.St.J.Farley and P.D.D'Eath, 
Phys Lett. B {\bf 613}, 181 (2005).\par           
\indent 
[3] A.N.St.J.Farley, 
'Quantum Amplitudes in Black-Hole Evaporation',           
Cambridge Ph.D. dissertation, approved 2002 (unpublished).\par
\in
[4] A.N.St.J.Farley and P.D.D'Eath,  
Class. Quantum Grav. {\bf 22}, 2765 (2005).\par
\in
[5] A.N.St.J.Farley and P.D.D'Eath, 
'Quantum Amplitudes in Black-Hole Evaporation: I. Complex Approach', 
submitted for publication (2005).\par
\in
[6] A.N.St.J.Farley and P.D.D'Eath, 
'Quantum Amplitudes in Black-Hole Evaporation: II. Spin-0 Amplitude', 
submitted for publication (2005).\par
\in      
[7] A.N.St.J.Farley and P.D.D'Eath, 
'Vaidya Space-Time in Black-Hole Evaporation', 
submitted for publication (2005).\par
\in 
[8] A.N.St.J.Farley and P.D.D'Eath, 
'Spin-1 Amplitudes in Black-Hole Evaporation', 
submitted for publication (2005).\par
\in
[9] A.N.St.J.Farley and P.D.D'Eath, 
'Spin-2 Amplitudes in Black-Hole Evaporation', 
submitted for publication (2005).\par
\in
[10] P.R.Garabedian, {\it Partial Differential Equations}, 
(Wiley, New York) (1964).\par
\indent 
[11] P.D.D'Eath, {\it Supersymmetric Quantum Cosmology}, 
(Cambridge University Press, Cambridge) (1996).\par
\indent
[12] R.P.Feynman and A.R.Hibbs, 
{\it Quantum Mechanics and Path Integrals}, 
(McGraw-Hill, New York) (1965).\par
\indent
[13] W.McLean, 
{\it Strongly Elliptic Systems and Boundary Integral Equations}, 
(Cambridge University Press, Cambridge) (2000); 
O.Reula, `A configuration space for quantum gravity and
solutions to the Euclidean Einstein equations in a slab region', 
Max-Planck-Institut f\"ur Astrophysik, {\bf MPA}, 275 (1987).\par
\indent
[14] J.Wess and J.Bagger, {\it Supersymmetry and Supergravity},
2nd. edition, (Princeton University Press, Princeton, N.J.) (1992).\par
\indent
[15]  P.D.D'Eath, 
'What local supersymmetry can do for quantum cosmology', 
in {\it The Future of Theoretical Physics and Cosmology}, 
eds. G.W.Gibbons, E.P.S.Shellard and S.J.Rankin 
(Cambridge University Press, Cambridge) 693 (2003).\par
\indent
[16] P.D.D'Eath, 
'Loop amplitudes in supergravity by canonical quantization', in 
{\it Fundamental Problems in Classical, Quantum and String Gravity}, 
ed. N. ${\rm S\acute a nchez}$ 
(Observatoire de Paris) {\bf 166} (1999), hep-th/9807028.\par
\indent
[17] A.Das, M.Fischler and M. Ro\v cek, 
Phys. Lett. B {\bf 69}, 186 (1977).\par
\indent
[18] L.D.Faddeev, in {\it Methods in Field Theory}, 
ed. R.Balian and J.Zinn-Justin (North-Holland, Amsterdam) (1976).\par
\indent
[19] L.D.Faddeev and A.A.Slavnov, 
{\it Gauge Fields: Introduction to Quantum Theory} 
(Benjamin/Cummings, Reading, Mass.) (1980).\par
\in
[20] C.Itzykson and J.-B.Zuber, {\it Quantum Field Theory},
(McGraw-Hill, New York) (1980).\par
\indent
[21] D.Bao, Y.Choquet-Bruhat, J.Isenberg and P.B.Yasskin, 
J. Math. Phys. {\bf 26}, 329 (1985).\par
\indent
[22] R.Penrose and W.Rindler, {\it Spinors and Space-Time}, vol. 1
(Cambridge University Press, Cambridge) (1984).\par
\in
[23] R.Penrose and W.Rindler, {\it Spinors and Space-Time}, vol. 2
(Cambridge University Press, Cambridge) (1986).\par
\in
[24] P.Breitenlohner and D.Z.Freedman, 
Phys. Lett. B {\bf 115}, 197 (1982).\par
\indent
[25] P.Breitenlohner and D.Z.Freedman, 
Ann. Phys. (N.Y.) {\bf 144}, 249 (1982).\par
\indent
[26] S.W.Hawking, Phys. Lett. B {\bf 126}, 175 (1983).\par
\indent
[27] P.C.Vaidya,  
Proc. Indian Acad. Sci. {\bf A33}, 264 (1951); R.W.Lindquist,
R.A.Schwartz and C.W. Misner, Phys. Rev. {\bf 137}, 1364 (1965).\par 
\indent
[28] F.A.Berezin, {\it The Method of Second Quantization}, 
(Academic, New York) (1966).\par
\indent
[29] P.D.D'Eath and G.V.M.Esposito, 
Phys. Rev. D {\bf 43}, 3234 (1991).\par
\in 
[30] P.D.D'Eath and G.V.M.Esposito, 
Phys. Rev. D {\bf 44}, 1713 (1991).\par
\in
[31] M.F.Atiyah, V.K.Patodi and I.M.Singer, 
Math. Proc. Cambridge Philos. Soc. {\bf 77}, 43 (1975).\par
\in
[32] S.A.Teukolsky, Astrophys. J. {\bf 185}, 635 (1973).\par
\in
[33] W.G.Unruh, Phys. Rev. Lett. {\bf 31}, 1265 (1973).\par
\in
[34] E.T.Newman and R.Penrose, J. Math. Phys. {\bf 3}, 566 (1962).\par
\indent
[35] W.Kinnersley, J. Math. Phys {\bf 10}, 1195 (1969).\par
\indent
[36] J.N.Goldberg, A.J.MacFarlane, E.T.Newman, F.Rohrlich and 
E.C.G.Sudarshan, J. Math. Phys. {\bf 8}, 2155 (1967).\par
\indent
[37] J.A.H.Futterman, F.A.Handler and R.A.Matzner,   
{\it Scattering from Black Holes}, 
(Cambridge University Press, Cambridge) (1988).\par
\indent 
[38]  M.Abramowitz and I.A.Stegun, 
{\it Handbook of Mathematical Functions}, (Dover, New York) (1964).\par
\indent
[39] T.Regge and J.A.Wheeler, Phys. Rev. {\bf 108}, 1063 (1957).\par
\indent
[40] C.W.Misner, K.S.Thorne and J.A.Wheeler, {\it Gravitation},
(Freeman, San Francisco) (1973).\par
\indent 
[41] G.V.M.Esposito, 
{\it Quantum Gravity, Quantum Cosmology and Lorentzian Geometries}, 
Lecture Notes in Physics m12 (Springer, Berlin) (1994).\par
\indent
[42]  G.V.M.Esposito, {\it Dirac Operators and Spectral Geometry},
(Cambridge University Press, Cambridge) (1998).\par
\indent
[43] A.N.St.J.Farley and P.D.D'Eath, 
'Spin-3/2 Amplitudes in Black-Hole Evaporation', in progress.\par

\end